# A comparison study on the growth pattern of traffic oscillations in car-following experiments

Shi-Teng Zheng[a], Rui Jiang[a,*], Junfang Tian[b], Xiaopeng Li[c], Bin Jia[a,*], Ziyou Gao[a], Shaowei Yu[d]

[a] *Key Laboratory of Transport Industry of Big Data Application Technologies for Comprehensive Transport, Ministry of Transport, Beijing Jiaotong University, Beijing 100044, China*

[b] *Institute of Systems Engineering, College of Management and Economics, Tianjin University, No. 92 Weijin Road, Nankai District, Tianjin 300072, China*

[c] *Department of Civil and Environmental Engineering, University of South Florida, FL 33620, USA*

[d] *School of Information Engineering, Chang'an University, Xi'an 710064, Shaanxi, China*

## ABSTRACT

The evolution of oscillations is a very important issue in traffic flow studies. A recent car-following experiment (Experiment-I) showed that the speed standard deviation grows in a concave way along a platoon of vehicles following one another. This finding indicates that the traditional traffic instability mechanism is debatable, in which the speed standard deviation initially grows in a convex way. This paper has investigated the growth pattern of traffic oscillations in another car-following experiment (Experiment-II) and compared it with that in Experiment-I. It is shown that the speed standard deviation also exhibits concave growth characteristics in Experiment-II. The paired-sample *t*-test and the Mann-Kendall (MK) trend test showed that there is no significant difference between the two datasets. However, the acceleration standard deviation was remarkably larger in Experiment-II since drivers were asked to follow closely. Furthermore, a comparison experiment has been performed which indicates that the set of experiments on a circular track can be considered equivalent to that on a straight track. Our study is expected to shed light not only on traffic flow dynamics itself but also on the future design of the experiment scheme.

## 1. Introduction

The traffic oscillation is an important feature of traffic flow. It refers to the phenomenon that instead of maintaining a comfortable and stable driving status, vehicles are often forced to repeatedly accelerate and decelerate (Li et al., 2014). Traffic oscillations usually form at the bottlenecks, such as highway lane drops (Bertini and Leal, 2005) and lane changes near merges and diverges (Mauch and Cassidy, 2002; Cassidy and Rudjanakanoknad, 2005; Laval and Daganzo, 2006; Laval et al., 2007; Ahn and Cassidy, 2007; Ahn et al., 2010; Zheng et al., 2011). Recent field experiments reveal that traffic oscillations can also arise even in the absence of bottlenecks and lane changing (Sugiyama et al., 2008; Stern et al., 2018). Traffic oscillation, as "a nuisance for all motorists throughout the world", is undesirable (Laval and Leclercq, 2010), leading to traffic breakdown and capacity drop (Chen et al., 2014; Yuan et al., 2017; Arnesen and Hjelkrema, 2018), driving discomfort, extra fuel consumption and emission, and more safety risks.

To quantify oscillation characteristics and identify contributing factors, many empirical studies have been performed, which provide solid evidences for periodical patterns of traffic oscillations in congested traffic (Koshi et al., 1981; Kühne, 1987; Ferrari, 1989; Mauch and Cassidy, 2002; Ahn, 2005; Schönhof and Helbing, 2007; Zielke et al., 2008). Mauch and Cassidy (2002) observed freeway traffic over multiple days on the 10-kilometer stretch of the Queen Elizabeth Way. Its findings indicated that oscillations arose only in queues, increased in amplitude and grew to their full amplitudes while propagating upstream. Laval et al. (2009) pointed out that oscillations had typical periods to be somewhere between 2 and 15 min. Gartner et al. (1992) studied the relationship between amplitude and period, and found that low frequencies are accompanied by high amplitudes and short periods result in small amplitudes. Mauch and Cassidy (2002) found that oscillations propagated at a nearly constant speed of about 22 to 24 km/h in Canada. Zielke et al. (2008) conducted a country-specific analysis of freeway traffic oscillations and pointed out that oscillations propagated at speeds of about 19 to 20 km/h on OR-217 in the United States, 16 km/h on the A9 in Germany, and 14 km/h on the M4 in the United Kingdom. Zheng et al. (2011) found that regardless of the trigger, the features of oscillation propagations were similar in terms of propagation speed, oscillation duration, and amplitude.

Inspired by the empirical observation findings, theoretical analyses and researches have been carried out to investigate the formation and propagation mechanisms of oscillations. In classical theoretical models, such as the General-Motors family of car-following models (Chandler et al., 1958; Gazis et al., 1959, 1961), the Payne model (Payne, 1979), the optimal velocity (OV) model (Bando et al., 1995), the intelligent driver model (IDM) (Treiber et al., 2000), and the full velocity difference (FVD) model (Jiang et al., 2001), the formation and propagation of traffic oscillations is due to the instability of steady state, which can be triggered by small perturbations. Li and Ouyang (2011) and Li et al. (2012, 2014) showed that nonlinear driving behavior might contribute to concave growth of traffic oscillations after the oscillation amplitude becomes significant.

Based on the analysis of detailed vehicle-trajectory data, Laval and Leclercq (2010) concluded that timid and aggressive driver behaviors are the cause for the spontaneous appearance of oscillations and their ensuing transformation into stop-and-go waves. They introduced a modified Newell model that allows vehicle trajectories to deviate from Newell trajectories to simulate the traffic oscillation. Chen et al. (2012a, 2012b) observed that the driving behavior before and after experiencing an oscillation might be different. They generalized the Laval-Leclercq model and proposed an "asymmetric behavioral model" to capture the behavior.

To study the growth pattern of oscillations, Jiang et al. (2014, 2015, 2018) and Huang et al. (2018) have conducted the car-following experiments in China. They found that the speed standard deviation increases in a concave way along the platoon. Tian et al. (2016) have investigated the growth pattern of traffic oscillations in the Next Generation Simulation (NGSIM) vehicle trajectory data. It was found that the speed standard deviation also grows concavely and is highly compatible with that in Chinese car-following experiments. They pointed out that in traditional car-following models, the speed standard deviation initially increases in a convex way, which is controversial to the experimental finding. They further argued that the mechanism of oscillation formation and propagation in traditional models is debatable. It was proposed that oscillations form and grow due to the cumulative effect of stochastic factors[1]. They also suggested that a model spanning a two-dimensional (2D) speed-spacing relation has the potential to reproduce the characteristics of traffic flow in experiments and empirical observations (Tian et al., 2017).

Repeatability is a necessary requirement of experimental studies. This paper aims to analyze whether the concave growth pattern of oscillations can be observed in other experiments and make a quantitative comparison. To this end, we analyze two other datasets collected from recent car-following experiments conducted in the USA (denoted as Experiment-II), which aims to study the effect of Cognitive and Autonomous Test (CAT) Vehicle on stop-and-go traffic. In Experiment-II, when the autonomous velocity controller or implemented human-executed dampening is activated, the CAT Vehicle actually tends to keep a larger gap from its preceding vehicle (or the last vehicle if we view the CAT Vehicle as the leader). In this way, the impact of the speed fluctuation of the preceding vehicle is mostly absorbed by this gap, and the CAT Vehicle can be viewed approximately as a leading vehicle cruising at a constant speed with little previous interferences, which is close to 25-car-platoon experiment[2] (denoted as Experiment-I). Therefore, we believe we make a fair comparison.

[1] We would like to mention that in several other works (Kim and Zhang, 2008; Yeo and Skabardonis, 2009; Laval et al., 2014), the nontrivial role of stochastic factors has also been studied.

This paper is organized as follows. Section 2 reviews the experimental setup of Experiment-II in details. Section 3 reports the comparison of the growth pattern of speed standard deviation between Experiment-I and Experiment-II. Section 4 further compares the growth pattern of acceleration standard deviation. Section 5 presents a new experimental study (denoted as Experiment-III) concerning to which extent the set of experiments on a circular track can be considered equivalent to that on a straight track. Finally, conclusions are given in Section 6.

## 2. Detailed description of data

In this section, we describe the vehicle velocity data collected from the field experiments in the USA. Experiment A, B, and C (Stern et al., 2018) were conducted on a single-lane circular track of 260-meter circumference with 21, 21 and 22 vehicles, respectively. In each experiment, there is only one *Cognitive and Autonomous Test* (CAT) Vehicle, which can be transitioned between manual velocity control and autonomous velocity control. The instruction provided to each driver prior to the start of the experiment is "Drive as if you were in rush hour traffic. Follow the vehicle ahead without falling behind. Do not pass the car ahead. Do not hit the car ahead. Drive safely at all times. Do not tailgate. But put an emphasis on catching up to the vehicle ahead, if a gap starts opening up." The data are illustrated in details as follows:

- Experiment A: 21 vehicles. The CAT Vehicle is initially under human control. After the first traffic wave is observed, the *FollowerStopper* controller[3] is activated with a desired velocity of 6.50 m/s. Then the velocity is changed to 7.00 m/s, 7.50 m/s, 8.00 m/s soon after, and is finally reduced to 7.50 m/s. After a while, the CAT Vehicle reverts to typical human driving behavior until the experiment is ended. See Stern et al. (2018) for the exact time of speed switching. The velocity profile of the CAT Vehicle and the evolution of the spatiotemporal pattern of car speed in Exp A are shown in Fig. 1.

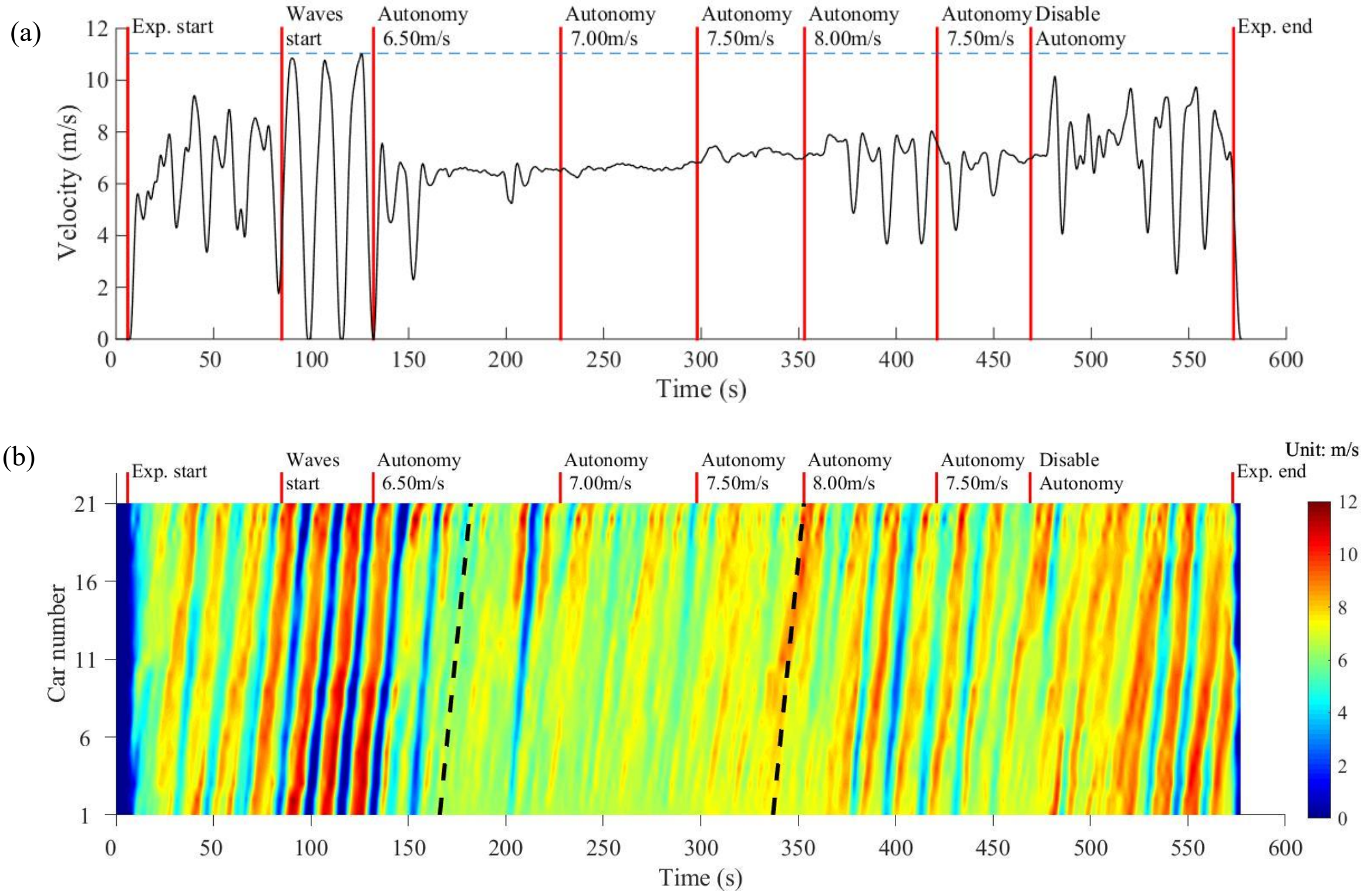

[2] The 25-car-platoon experiment was carried out on January 19, 2013 on a 3.2 km stretch of the Chuangxin Avenue in a suburban area in Hefei City, China. See Jiang et al. (2014, 2015) for more details.

[3] See section 3.1 in Stern et al. (2018) for more details about the *FollowerStopper* controller.

**Fig. 1.** (a) Velocity profile of the CAT Vehicle (the leading car) in Exp A. (b) Evolution of the spatiotemporal pattern of car speed in Exp A. For details on the two oblique black dotted lines, see section 3.1.

- Experiment B: 21 vehicles. The CAT Vehicle is always under human control, but after a wave initially appears, the driver of the CAT Vehicle switches from initially typical human driving behavior to the manual control strategy implemented by a trained human driver without the aid of automation. The CAT Vehicle is commanded at a desired velocity of 6.26 m/s at first, then increases to 7.15 m/s, and reverts to typical human driving behavior after a while until the experiment is ended. The velocity profile of the CAT Vehicle and the evolution of the spatiotemporal pattern of car speed in Exp B are shown in Fig. 2.

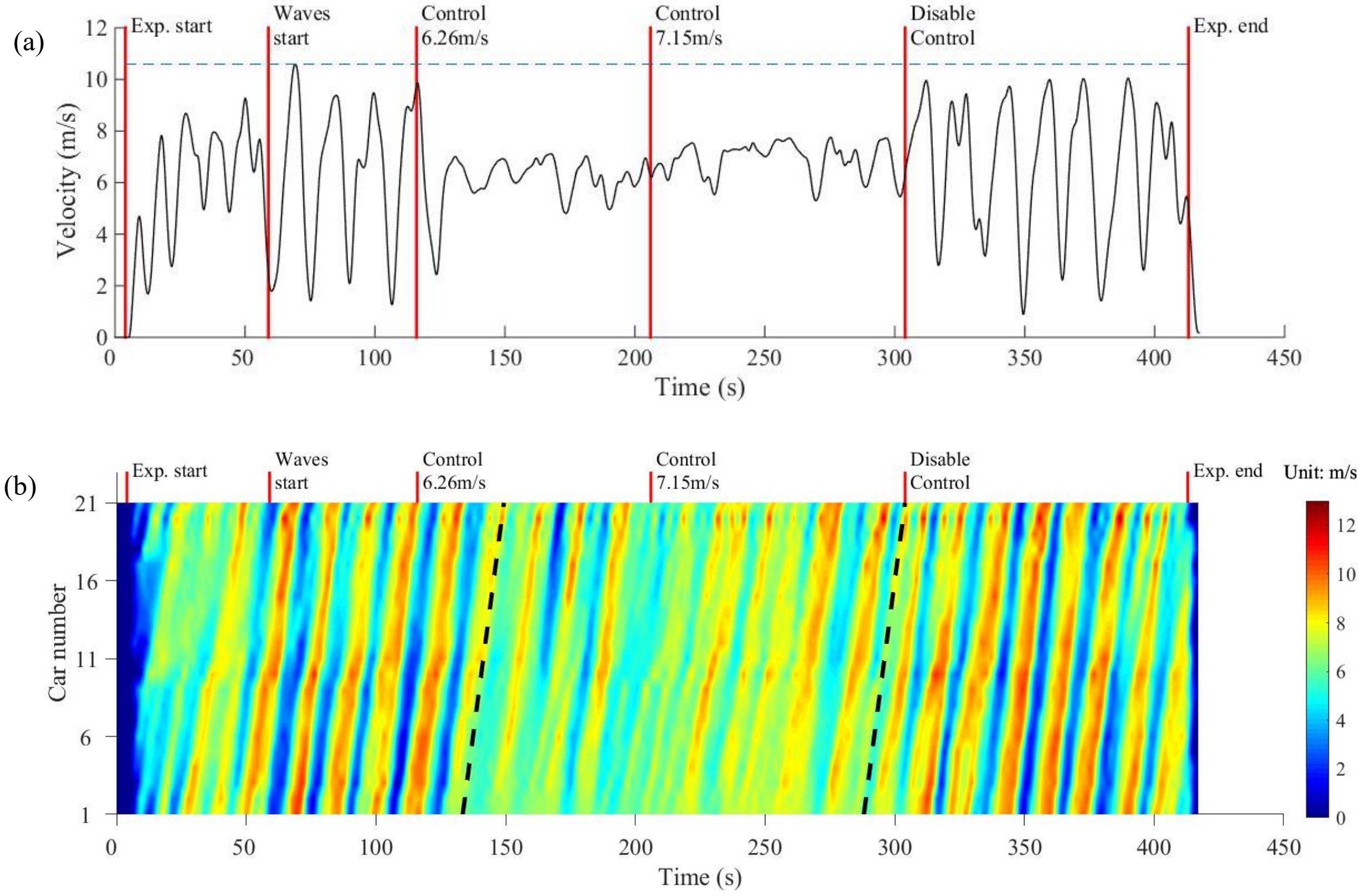

**Fig. 2.** (a) Velocity profile of the CAT Vehicle (the leading car) in Exp B. (b) Evolution of the spatiotemporal pattern of car speed in Exp B.

- Experiment C: 22 vehicles. The CAT Vehicle is initially under human control. After a wave initially appears, the *PI controller with saturation* wave damping controller[4] is activated, and remains active until the end of the experiment. The velocity profile of the CAT Vehicle and the evolution of the spatiotemporal pattern of car speed in Exp C are shown in Fig. 3.

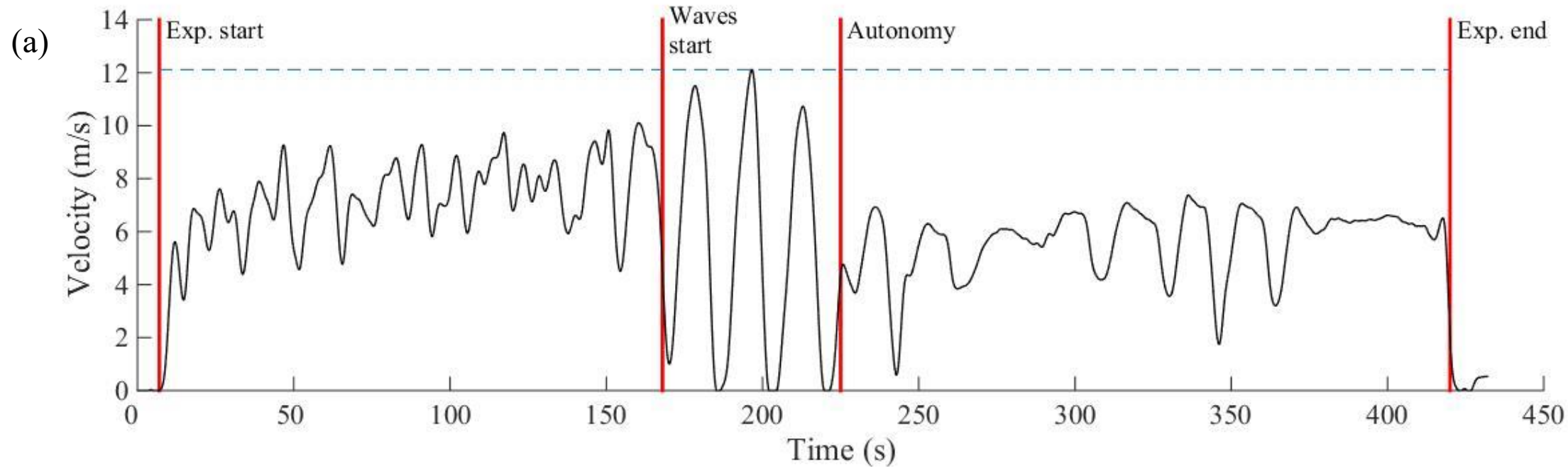

---

[4] See section 3.2 in Stern et al. (2018) for more details about the *PI with saturation* controller.

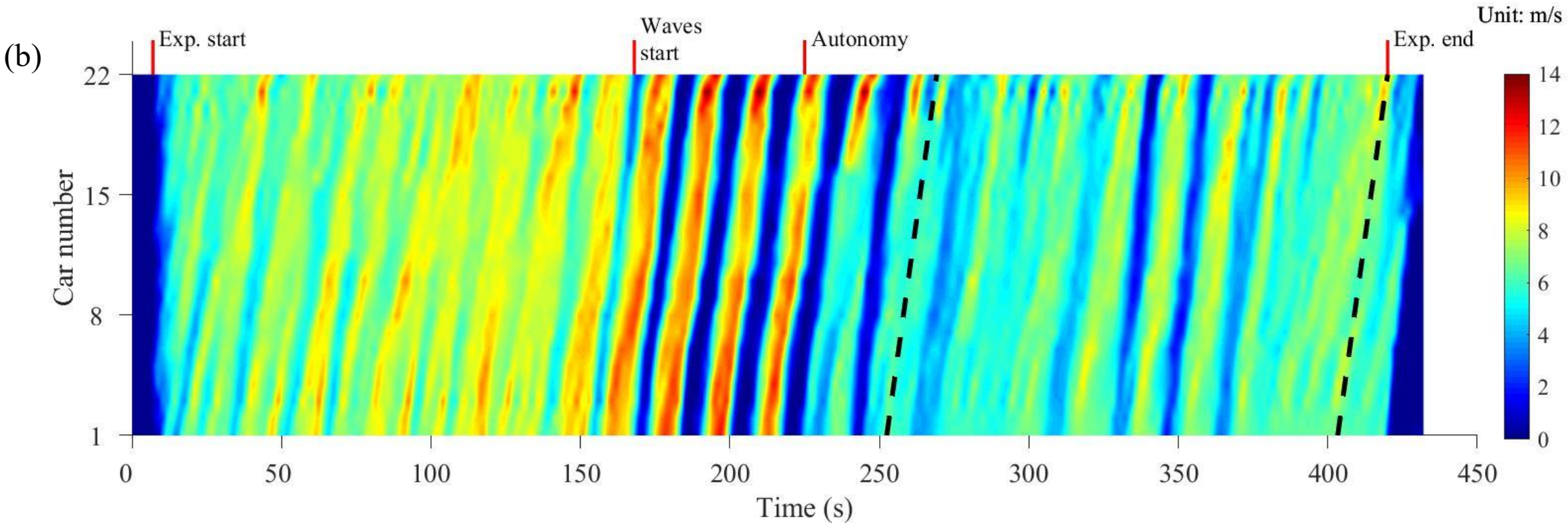

**Fig. 3.** (a) Velocity profile of the CAT Vehicle (the leading car) in Exp C. (b) Evolution of the spatiotemporal pattern of car speed in Exp C.

The Experiment F, G and H (Wu et al., 2019) were conducted on a single-lane circular track in 260 m circumference, the same experimental field as Exp A, B, and C, with 19, 21 and 22 vehicles respectively in Tucson, Arizona in July 2016. The drivers are instructed to "Safely follow the vehicle in front as if in rush hour traffic, but in addition place an emphasis on closing the gap with the vehicle in front, whenever safety permits." See Wu et al. (2019) for more details. The data of Experiment F, G and H are illustrated in details as follows:

- Experiment F: 19 vehicles. The CAT Vehicle is always under human control. About 60 seconds later, the driver of the CAT Vehicle is told via radio to slow down and maintain a constant speed. The velocity profile of the CAT Vehicle and the evolution of the spatiotemporal pattern of car speed in Exp F are shown in Fig. 4.

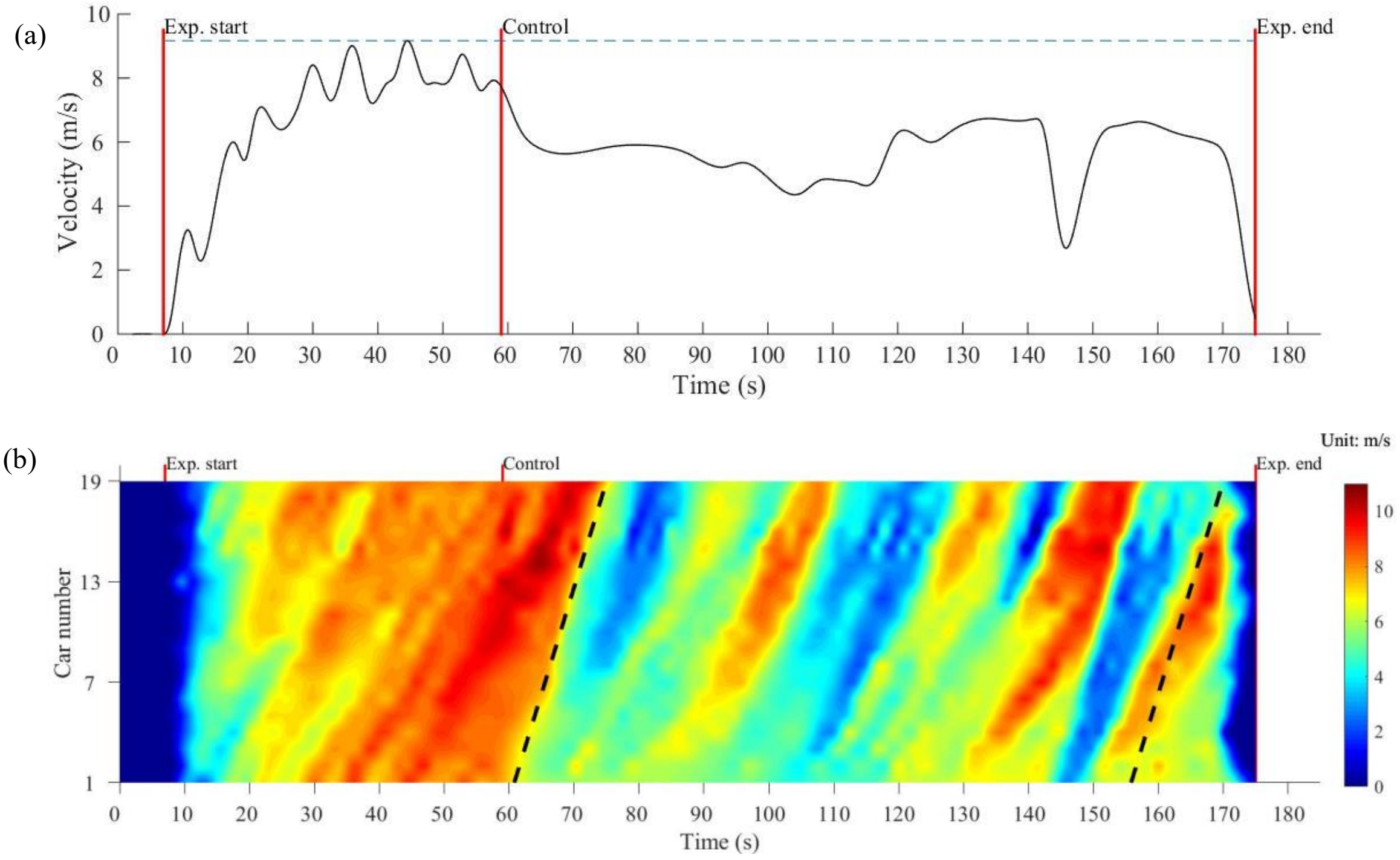

**Fig. 4.** (a) Velocity profile of the CAT Vehicle (the leading car) in Exp F. (b) Evolution of the spatiotemporal pattern of car speed in Exp F.

- Experiment G: 21 vehicles. The CAT Vehicle is always under human control. Similar to Exp F, the driver of the CAT Vehicle is instructed to drive with a constant speed of 6.26 m/s (14 mph). The velocity profile of the CAT Vehicle and the evolution of the spatiotemporal pattern of car speed in Exp G are shown in Fig. 5.

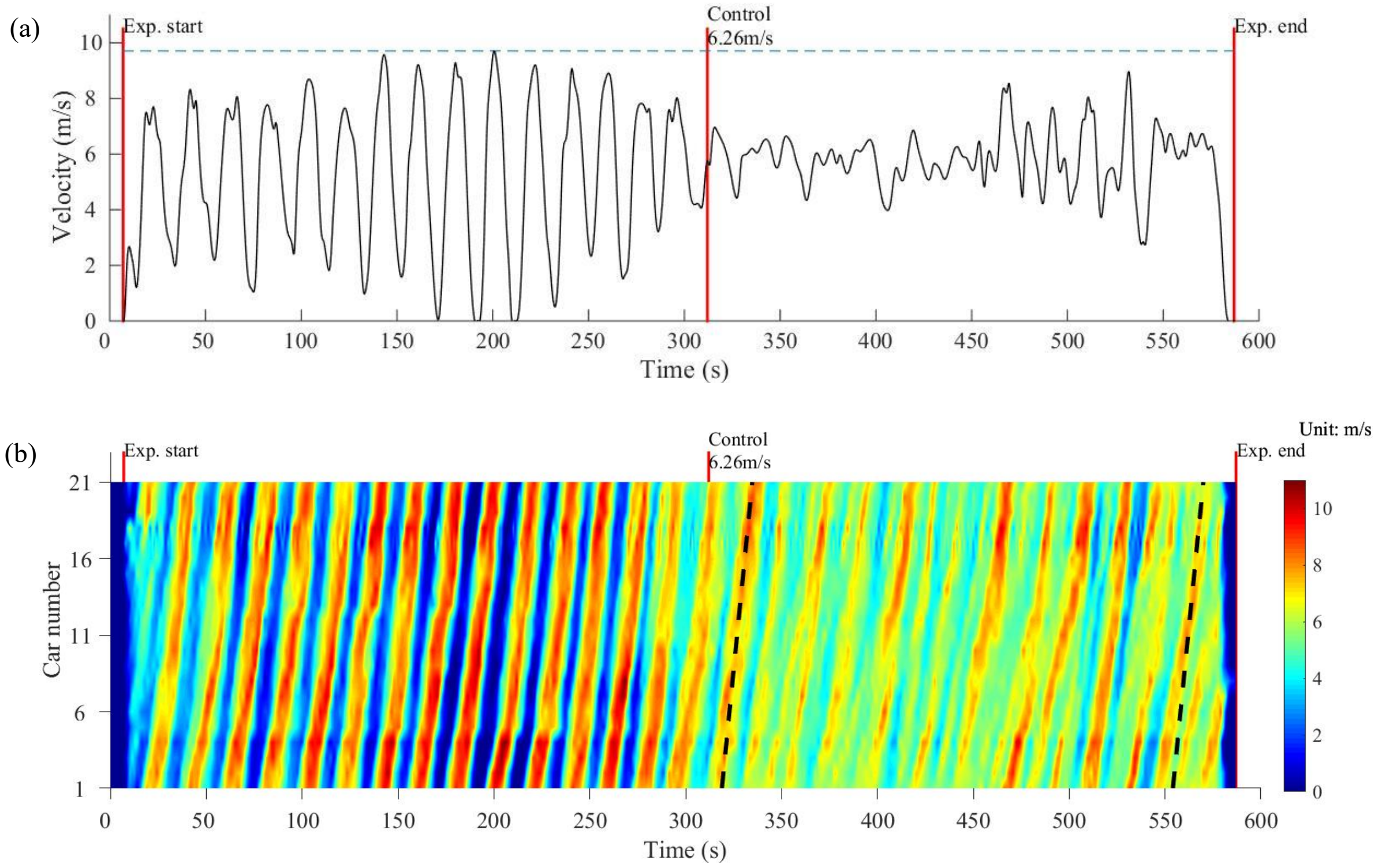

**Fig. 5.** (a) Velocity profile of the CAT Vehicle (the leading car) in Exp G. (b) Evolution of the spatiotemporal pattern of car speed in Exp G.

- Experiment H: 22 vehicles. Similar to Exp F and Exp G, the driver of the CAT Vehicle is instructed to maintain a constant speed for some time. In Exp H, this occurs twice: for the first time, the driver of the CAT Vehicle is instructed to drive at 5.36 m/s (12 mph) after 191 seconds and is instructed again to reduce the speed to 4.47 m/s (10 mph) after 411 seconds. The velocity profile of the CAT Vehicle and the evolution of the spatiotemporal pattern of car speed in Exp H are shown in Fig. 6.

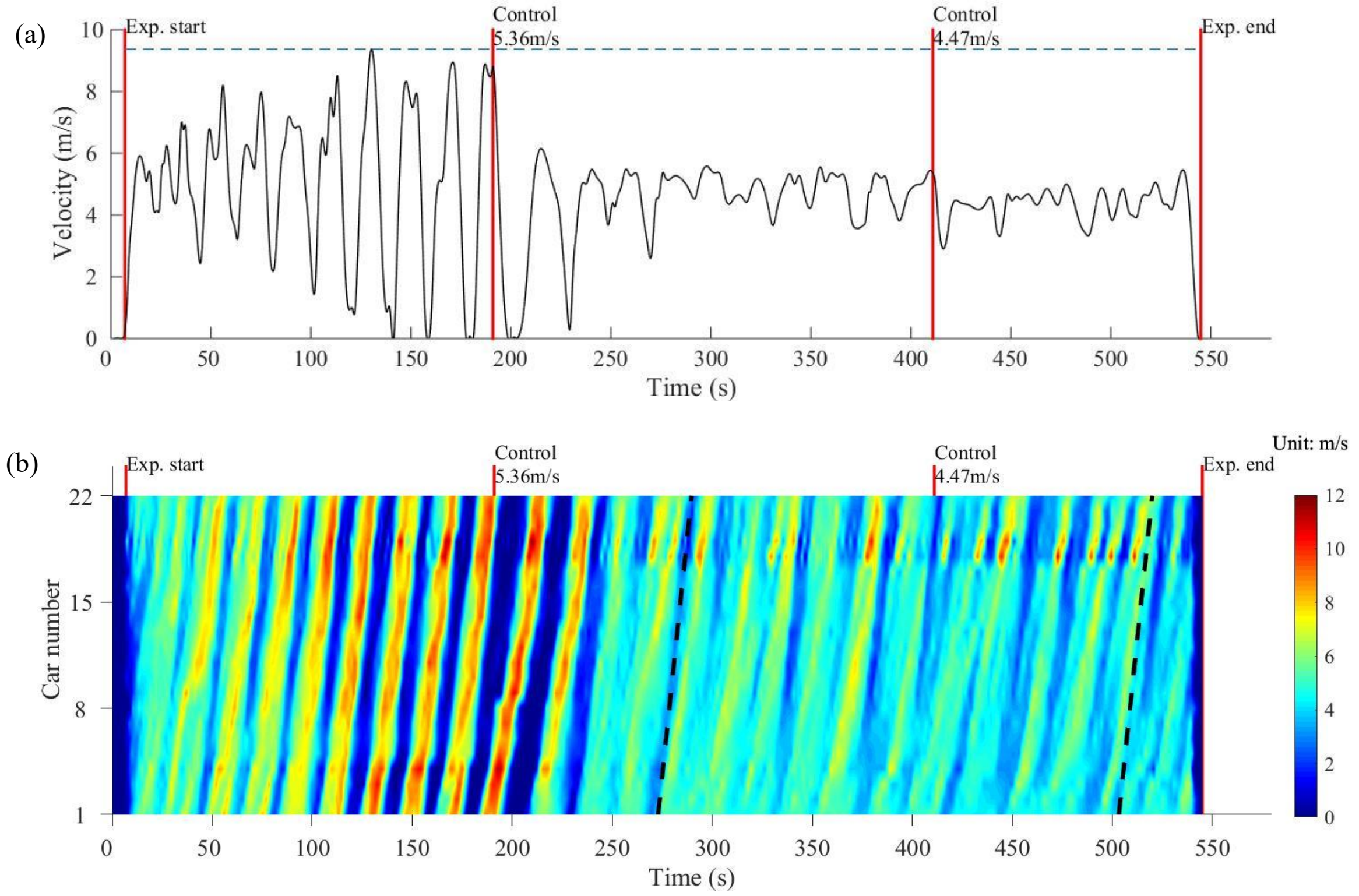

**Fig. 6.** (a) Velocity profile of the CAT Vehicle (the leading car) in Exp H. (b) Evolution of the spatiotemporal pattern of car speed in Exp H.

## 3. The growth pattern of traffic oscillations

### *3.1. The concerned interval in the Experiment-II*

Although Experiment-II is conducted on a circular track, the CAT Vehicle has been switched into autonomous velocity control or implemented human-executed control mode for a certain time interval, during which the driving behavior of the CAT Vehicle is no longer the typical human-driving behavior. It has a large distance from the preceding car and the amplitude of speed fluctuation becomes small. Thus, we consider these vehicles on the circular track as a platoon in this interval and regard the CAT Vehicle as the leading car of this platoon to study traffic oscillations.

Take Exp A as an example. From the two patterns of Exp A (Fig. 1), it is easy to figure out that the velocity becomes relatively stable after the autonomous velocity controller is activated, and the wave is noticeably affected after the 165th seconds. When the velocity of the CAT Vehicle is further increased to 8.00 m/s at the 347th seconds, a traffic wave rises again. Thus, we focus on the interval between the 165th seconds and the 347th seconds, during which the speed fluctuation is much smaller than other intervals. The interval we concerned is about 170 seconds, during which the average speed of the leading car is calculated to be 6.65 m/s. For the following vehicles, the corresponding intervals are indicated by two oblique black dotted lines considering the back-propagation velocity of traffic oscillations (see Fig. 1(b)).

The concerned intervals in other experiments are extracted similarly, see Fig. 2(b) ~ Fig. 6(b), respectively. The duration of the concerned interval and the average speed of the leading car for Exp A, B, C, F, G, and H are shown in Table 1. Since the average speed of the leading car in Exp A and B is close to 25 km/h, the average speed of the leading car in Exp C, F, and G is close to 20 km/h and that in Exp H is close to 15 km/h, we divide them into three groups. In the next section, we will compare the experimental average results in each group with those in Experiment-I at the leading car speed of 25 km/h, 20 km/h and 15 km/h, respectively.

**Table 1**

The duration of the concerned interval and the average speed of the leading car

| | Group 1 | | Group 2 | | | Group 3 |
|---|---|---|---|---|---|---|
| | Exp A | Exp B | Exp C | Exp F | Exp G | Exp H |
| Duration | ~ 170 s | ~ 155 s | ~ 150 s | ~ 95 s | ~ 235 s | ~ 230 s |
| Average speed of the leading car | 6.65 m/s (23.9 km/h) | 6.60 m/s (23.8 km/h) | 5.78 m/s (20.8 km/h) | 5.30 m/s (19.1 km/h) | 5.81 m/s (20.9 km/h) | 4.61 m/s (16.6 km/h) |

### *3.2. The growth pattern of speed standard deviation*

Stripe structure has been observed in the evolution of the spatiotemporal pattern of traffic flow, which exhibits the formation, propagation, growth and dissipation of oscillations (Fig. 1(b) ~ Fig. 6(b)). This looks similar to that observed in Chinese experiments. To investigate the growth pattern of traffic oscillations, we calculate the speed standard deviation of all vehicles of Exp A, B, C, F, G, and H, and compare the average results in each group with those in Experiment-I. The comparison results are presented in Fig. 7. One can see that the results of Experiment-II in each group accord well with those of Experiment-I at the similar speed of the leading car.

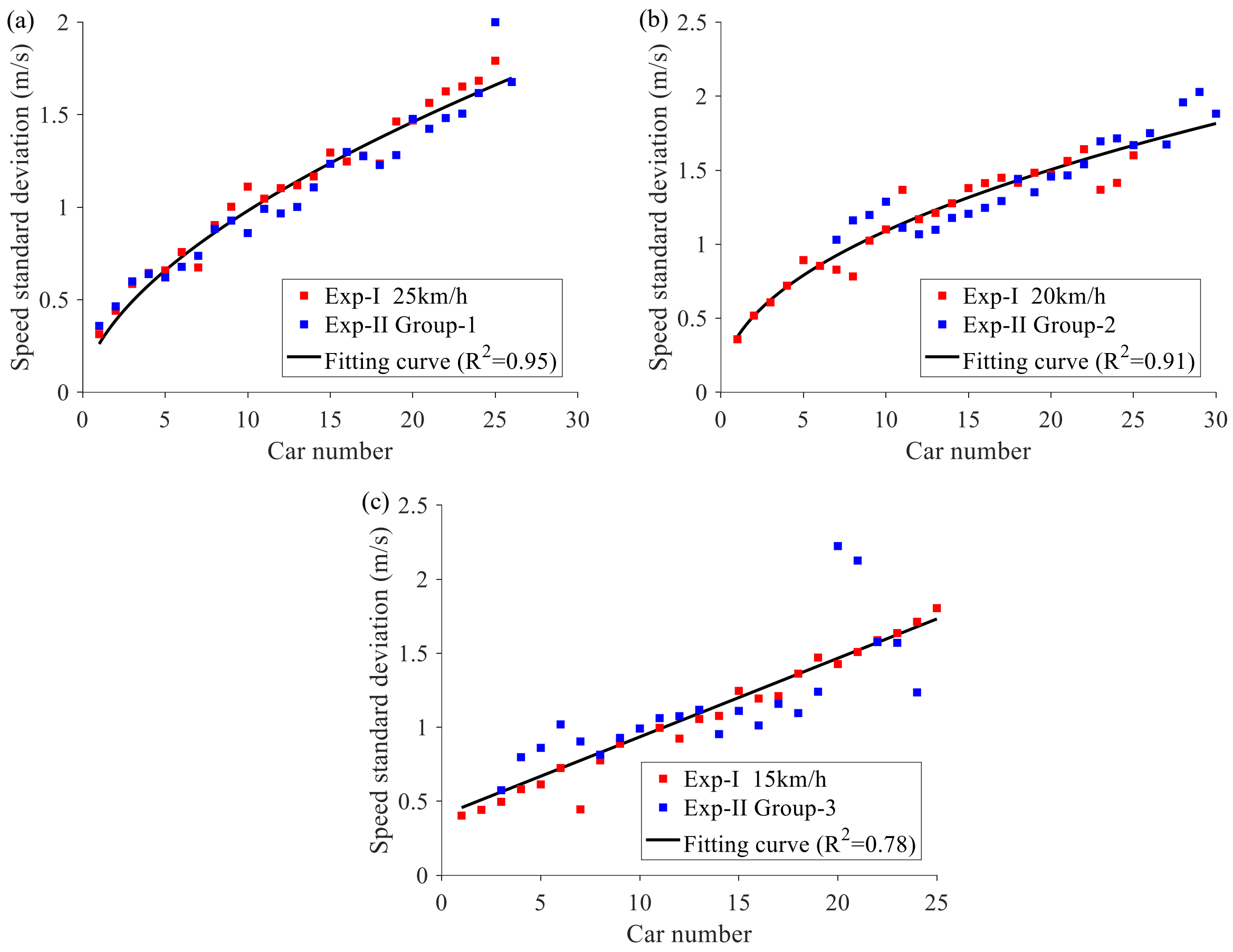

**Fig. 7.** Comparison of the speed standard deviation of all vehicles between Experiment-I and Experiment-II at the similar speed of the leading car. Note that the Experiment-II data have been shifted horizontally as proposed in Tian et al. (2016). The fitting curves are given by $f(x) = ax^b + c$ in (a) and (b), and $f(x) = ax + b$ in (c), where (a) $a$ = 0.2610, $b$ = 0.5748, $c$ = 0; (b) $a$ = 0.3743, $b$ = 0.4643, $c$ = 0; (c) $a$ = 0.0531, $b$ = 0.4041.

To compare the Experiment-I and Experiment-II results quantitatively, the paired-sample *t*-test and the Mann-Kendall trend test are applied to the two datasets at the 1% significance level. Table 2 shows the details of the two paired datasets of speed standard deviation used in the *t*-test. Their difference is used in the Mann-Kendall trend test. The *p*-value equals 0.0206, 0.7701, and 0.2780 in the *t*-test for the three pairs of data, respectively. The null hypothesis in the *t*-test that there is no significant difference in the speed standard deviation between Experiment-I and Experiment-II could not be rejected ($p$-value > 0.01). The *p*-value equals 0.0973, 0.8887, and 0.0131 in the Mann-Kendall trend test. The null hypothesis that no significant trend exists could not be rejected ($p$-value > 0.01). The two tests thus demonstrate that there is no significant difference between the two datasets concerning the growth of speed standard deviation.

One might notice that the growth of speed standard deviation has been fitted linearly in Fig.7(c). Actually, as pointed out in Jiang et al. (2015), "if we had a much longer platoon, the variations of speed of cars in the tail of the platoon would be capped and the line would bend downward, making the overall curve concave shaped." The 51-car-platoon experiment reported in Jiang et al. (2018) has validated this argument.

**Table 2**

Details of the two paired datasets of speed/acceleration standard deviation used in the hypothesis test (unit: m·s$^{-1}$/m·s$^{-2}$). "—" means that either Experiment-I data or Experiment-II data are lacked.

| Car number | The 1st pair (Fig.7/9(a)) | | The 2nd pair (Fig. 7/9(b)) | | The 3rd pair (Fig.7/9(c)) | | The 4th pair (Fig.13(a)/(b)) | |
|---|---|---|---|---|---|---|---|---|
| | Exp-I | Exp-II | Exp-I | Exp-II | Exp-I | Exp-II | Exp-III | Exp-III |

| | 25 km/h | Group-1 | 20 km/h | Group-2 | 15 km/h | Group-3 | Circular track | Straight track |
|---|---|---|---|---|---|---|---|---|
| 1 | 0.3140 /0.0943 | 0.3588 /0.2181 | — | — | — | — | 0.0557 /0.0751 | 0.0647 /0.0873 |
| 2 | 0.4408 /0.1588 | 0.4645 /0.3870 | — | — | — | — | 0.2676 /0.1514 | 0.2871 /0.1526 |
| 3 | 0.5858 /0.2317 | 0.5995 /0.6459 | — | — | 0.4966 /0.2283 | 0.5749 /0.2055 | 0.4430 /0.2122 | 0.4331 /0.2027 |
| 4 | 0.6451 /0.2309 | 0.6389 /0.6479 | — | — | 0.5822 /0.2559 | 0.7981 /0.4421 | 0.4891 /0.1699 | 0.4906 /0.1642 |
| 5 | 0.6588 /0.275 | 0.6207 /0.5237 | — | — | 0.6142 /0.3084 | 0.8612 /0.4555 | 0.7583 /0.2729 | 0.7901 /0.3089 |
| 6 | 0.7577 /0.2823 | 0.6780 /0.4142 | — | — | 0.7249 /0.3418 | 1.0195 /0.5078 | 0.8513 /0.3144 | 0.9320 /0.3745 |
| 7 | 0.6745 /0.1693 | 0.7374 /0.5385 | 0.8282 /0.176 | 1.0306 /0.4686 | 0.4455 /0.1029 | 0.9043 /0.3953 | 0.8407 /0.3073 | 0.8722 /0.3637 |
| 8 | 0.9037 /0.294 | 0.8837 /0.6181 | 0.7825 /0.264 | 1.1613 /0.6002 | 0.7766 /0.3556 | 0.8149 /0.3319 | 0.9439 /0.3507 | 0.9050 /0.4236 |
| 9 | 1.0026 /0.3202 | 0.9279 /0.6572 | 1.0244 /0.2799 | 1.1979 /0.5875 | 0.8891 /0.4548 | 0.9285 /0.4744 | 0.9325 /0.3966 | 0.9705 /0.3942 |
| 10 | 1.1114 /0.3952 | 0.8605 /0.5361 | 1.1003 /0.3009 | 1.2874 /0.7036 | 0.9913 /0.4486 | 0.9923 /0.4495 | 0.9735 /0.3552 | 1.0313 /0.3442 |
| 11 | 1.0465 /0.3609 | 0.9913 /0.5886 | 1.3678 /0.3020 | 1.1110 /0.664 | 0.9965 /0.4771 | 1.0620 /0.5295 | 1.0410 /0.4445 | 1.1758 /0.5054 |
| 12 | 1.1032 /0.3176 | 0.9666 /0.5421 | 1.1690 /0.277 | 1.0676 /0.5946 | 0.9239 /0.3531 | 1.0753 /0.4482 | 1.0665 /0.4774 | 1.2948 /0.6030 |
| 13 | 1.1190 /0.3434 | 1.0016 /0.559 | 1.2116 /0.2891 | 1.0977 /0.6067 | 1.0561 /0.4243 | 1.1184 /0.5547 | — | — |
| 14 | 1.1665 /0.352 | 1.1076 /0.6767 | 1.2768 /0.2818 | 1.1780 /0.5928 | 1.0770 /0.4141 | 0.9540 /0.4774 | — | — |
| 15 | 1.2959 /0.3666 | 1.2348 /0.7737 | 1.3794 /0.3024 | 1.2056 /0.5866 | 1.2460 /0.4026 | 1.1108 /0.5541 | — | — |
| 16 | 1.2473 /0.3554 | 1.2987 /0.7252 | 1.4131 /0.3291 | 1.246 /0.569 | 1.1944 /0.3771 | 1.0125 /0.4318 | — | — |
| 17 | 1.2798 /0.3967 | 1.2758 /0.6976 | 1.4492 /0.3885 | 1.2918 /0.5838 | 1.2107 /0.4201 | 1.1585 /0.5618 | — | — |
| 18 | 1.2358 /0.2862 | 1.2282 /0.6401 | 1.4154 /0.2837 | 1.4414 /0.6667 | 1.3624 /0.4203 | 1.0959 /0.5141 | — | — |
| 19 | 1.4635 /0.4039 | 1.2818 /0.8493 | 1.4832 /0.3046 | 1.3513 /0.5955 | 1.4711 /0.4705 | 1.2403 /0.7074 | — | — |
| 20 | 1.4692 /0.4155 | 1.4775 /1.1117 | 1.4718 /0.3562 | 1.4572 /0.6238 | 1.4272 /0.4827 | 2.2243 /1.2901 | — | — |
| 21 | 1.5641 /0.4583 | 1.4240 /0.8649 | 1.5627 /0.3618 | 1.4652 /0.6297 | 1.5088 /0.4966 | 2.1264 /1.2655 | — | — |
| 22 | 1.6262 /0.4635 | 1.4821 /0.9495 | 1.6422 /0.3879 | 1.5401 /0.6681 | 1.5887 /0.5247 | 1.5758 /0.7024 | — | — |
| 23 | 1.6526 /0.4155 | 1.5059 /0.8112 | 1.3681 /0.2656 | 1.6952 /0.8916 | 1.6365 /0.4728 | 1.5708 /0.7623 | — | — |
| 24 | 1.6837 /0.4629 | 1.6172 /1.1151 | 1.4148 /0.3048 | 1.7153 /0.8381 | 1.7129 /0.5644 | 1.2358 /0.4762 | — | — |
| 25 | 1.7919 /0.5149 | 2.0001 /1.7741 | 1.6008 /0.3792 | 1.6698 /0.7458 | — | — | — | — |

## 4. The growth pattern of acceleration standard deviation

The speed standard deviation measures the amplitude of speed oscillation, which is not enough to characterize the speed oscillation. Another important feature of speed oscillation is the frequency. The acceleration standard deviation can, to some extent, reflect the frequency of speed oscillation. Actually, if the speed time series is a simple periodic sinusoid wave, denoted as $v_n(t) = A\sin(2\pi f \cdot t) + b$, then its acceleration $a_n(t) = 2\pi f \cdot A\cos(2\pi f \cdot t)$. The standard deviations of speed time series and acceleration time series are $(\sqrt{2}A)/2$ and $\sqrt{2}A \cdot \pi f$, respectively. The acceleration standard deviation is thus proportional to the frequency *f*.

To calculate the acceleration, the speed signals are firstly smoothed with a low-pass Fourier filter with a cut-off frequency of 0.6 Hz[5]. An example of the smoothed speed time series of Experiment-I and Experiment-II (red color line) is shown in Fig. 8(a) and (b), respectively. One can see that the random fluctuations are essentially eliminated.

Now we calculate the acceleration via

$$a_n(t) = \frac{v_n(t) - v_n(t - \Delta t)}{\Delta t} \tag{1}$$

where $a_n(t)$ and $v_n(t)$ are the acceleration and speed of vehicle $n$ at time $t$, respectively; $\Delta t$ = 0.1 s. The corresponding acceleration time series of Experiment-I and Experiment-II are shown in Fig. 8(c) and (d), respectively. One can see that the acceleration noises are also eliminated.

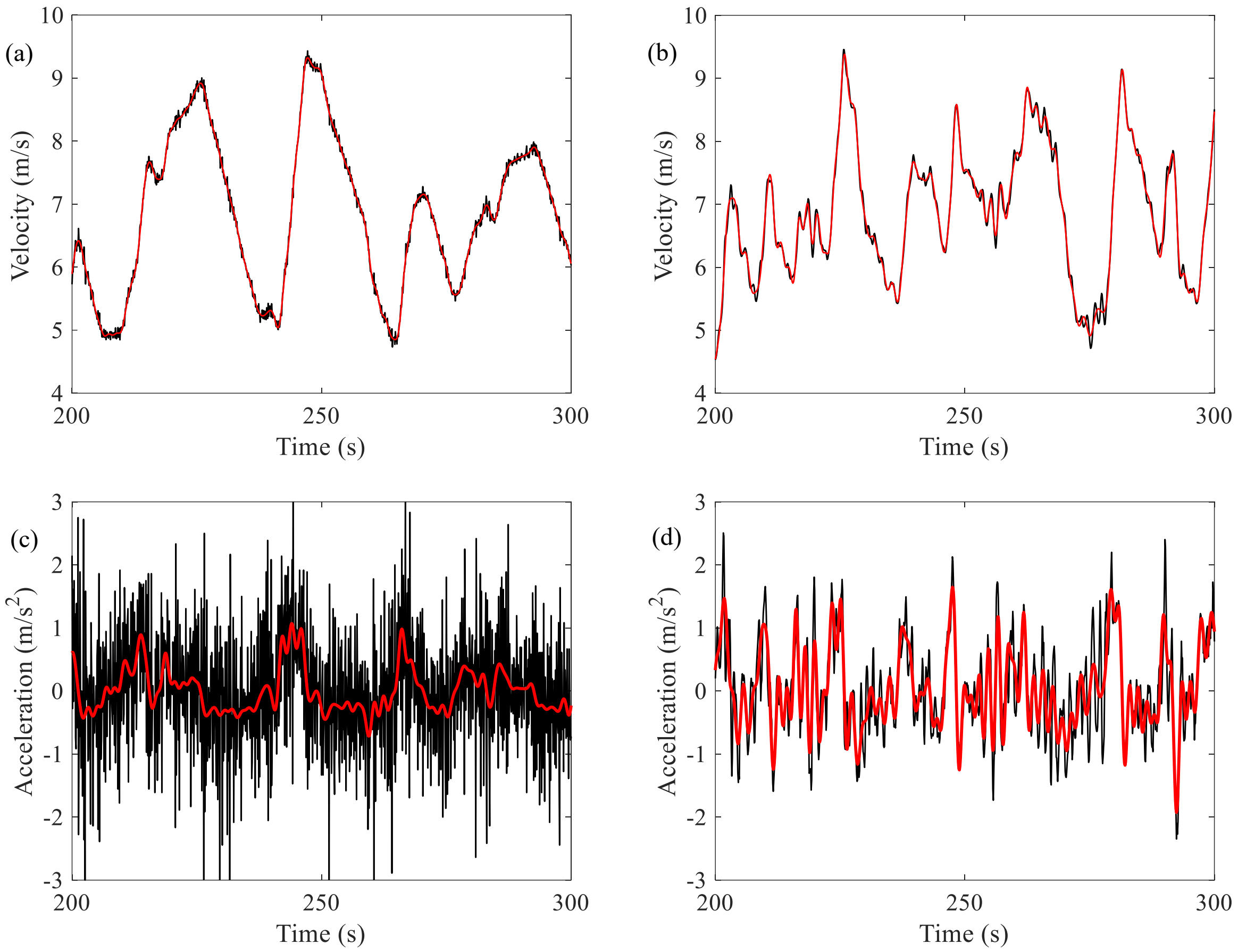

**Fig. 8.** (a) The speed of the 13th vehicle in one run in Experiment-I, and (b) the speed of the 8th vehicle in Exp B of Experiment-II. Red line is the smoothed speed signal, while black line is the raw one. (c), (d) The acceleration of the vehicle in (a), (b), respectively. Red line is calculated from the smoothed speed signal, while black line is calculated from the raw one.

From Fig. 8, it can be seen that speed oscillation amplitude of the two vehicles is roughly equal. However, speed oscillation frequency of the vehicle in Experiment-II is remarkably larger than that in Experiment-I. As a result, the vehicle in Experiment-II has a larger acceleration standard deviation.

Fig. 9 shows the growth pattern of acceleration standard deviation. One can see that the patterns are significantly different between Experiment-I and Experiment-II: the data in Experiment-II are larger than ones in Experiment-I. We also test the similarity of acceleration standard deviation between Experiment-I and Experiment-II. The two paired datasets of acceleration standard deviation used in the *t*-test are shown in Table 2. The *p*-value in the *t*-test equals 1.9293e-08, 2.0618e-11, and 0.0018 for the three pairs of data in Fig. 9, respectively. The null hypothesis in the *t*-test that there is no significant difference in the acceleration standard deviation between Experiment-I and Experiment-II is rejected ($p$-value < 0.01).

[5] We have examined the results by changing the value in the range between 0.3 Hz and 1 Hz, and found only minor quantitative differences.

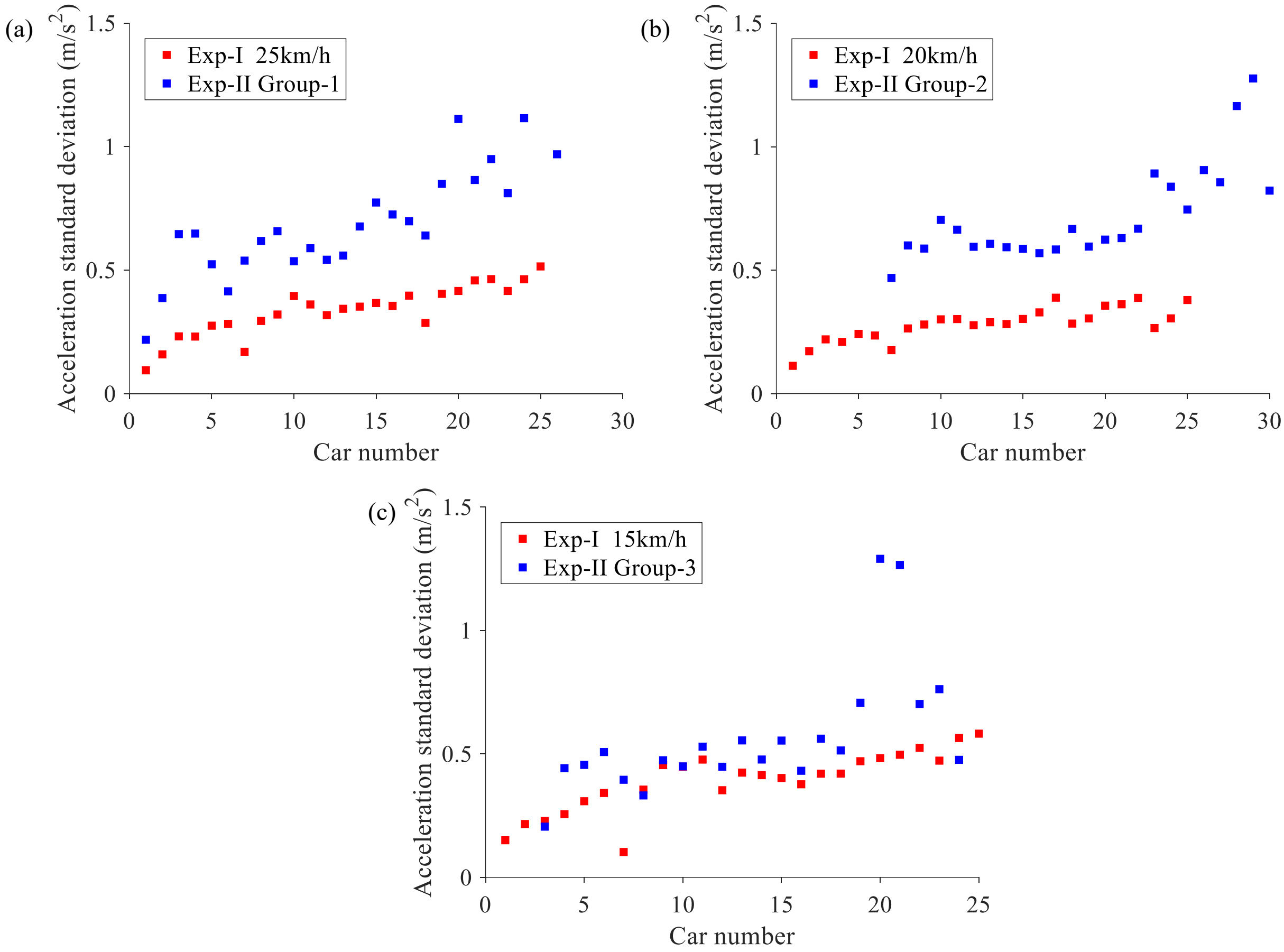

**Fig. 9.** Comparison of the acceleration standard deviation of all vehicles between Experiment-I and Experiment-II at the similar speed of the leading car. Note that the data in Experiment-II have been shifted horizontally as in Fig. 7.

There are two possible reasons for the different growth pattern of acceleration standard deviation between Experiment-I and Experiment-II: (i) Experiment-I and Experiment-II are carried out on straight and circular tracks respectively; (ii) the experimental instructions are different. In the next section, we reported a new comparison experiment, which demonstrates that experiments on a circular track can be considered equivalent to that on a straight track.

Thus, next we consider reason (ii). To this end, we compare the average distance headway. Fig. 10 shows that at leading speed of 20 km/h or 25 km/h, the average distance headways in Experiment-II are all notably smaller than those in Experiment-I. This is because, in Experiment-II, the instruction for drivers is “Drive as if you were in rush hour traffic. And put an emphasis on catching up to the vehicle ahead.”, while it is “Drive normally” in Experiment-I. Due to the instruction, the drivers follow more closely in Experiment-II. Consequently, speed oscillation frequency, and thus acceleration standard deviation in Experiment-II is larger than that in Experiment-I.

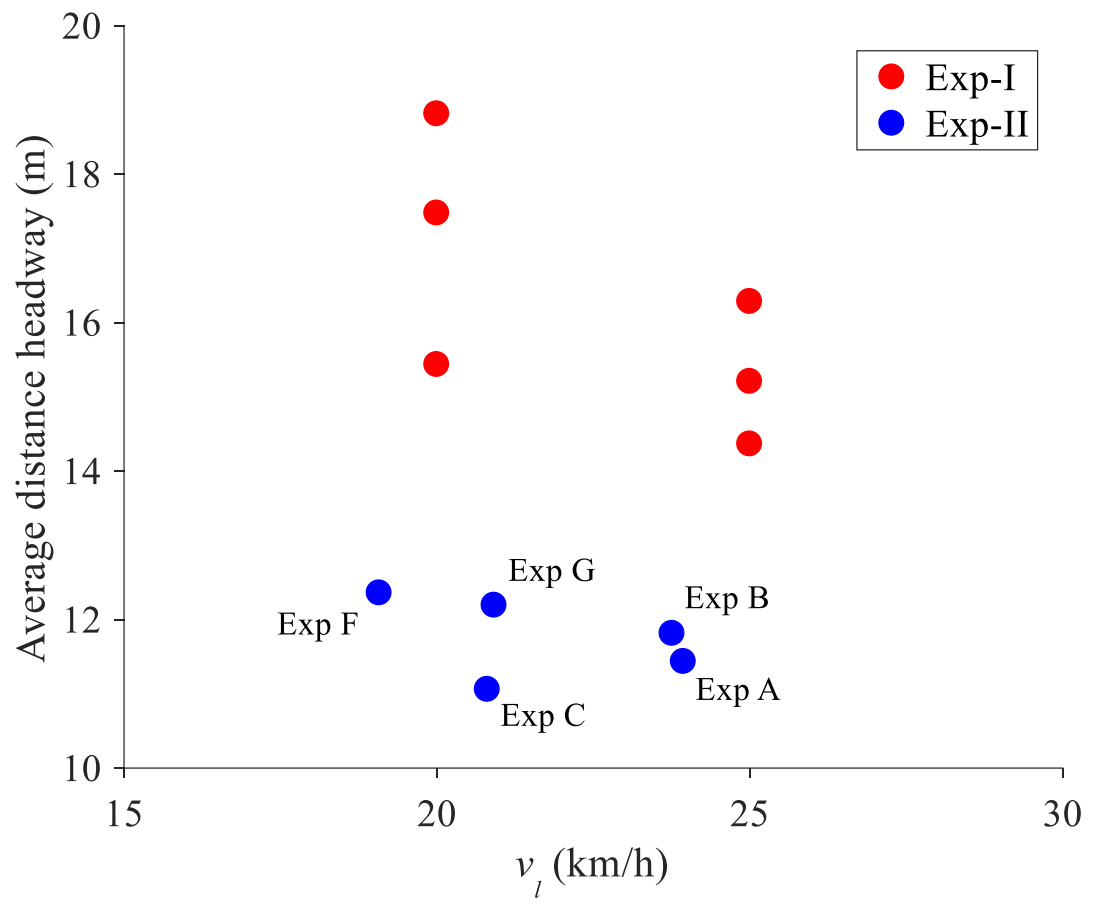

**Fig. 10.** The comparison of the average distance headway between Experiment-I and Experiment-II. Each data point corresponds to one run of experiment. $v_l$ denotes the average velocity of the leading car.

## 5. Straight track vs. circular track: an experimental study

The Experiment-I was conducted on a straight road while the Experiment-II used a quite small ring. To which extent these two sets of experiments can be considered equivalent? To explore the issue, we performed an experimental study for comparison. The experiment (Experiment-III) was carried out on December 1, 2018 in Automobile Experiment Base of Chang'an University, China. We setup a single-lane circular track on the automobile test square using traffic cones. The inner diameter is 78 m and the outer diameter is 82 m. Thus, the track width is 4 m and the effective diameter is 80 m, and the effective circumference is about 251 meters, which is close to that in Experiment-II. We also use the automobile test track, which consists of two 665-meter straight tracks and two bends at both ends. See Fig. 11 for the sketch of the two tracks.

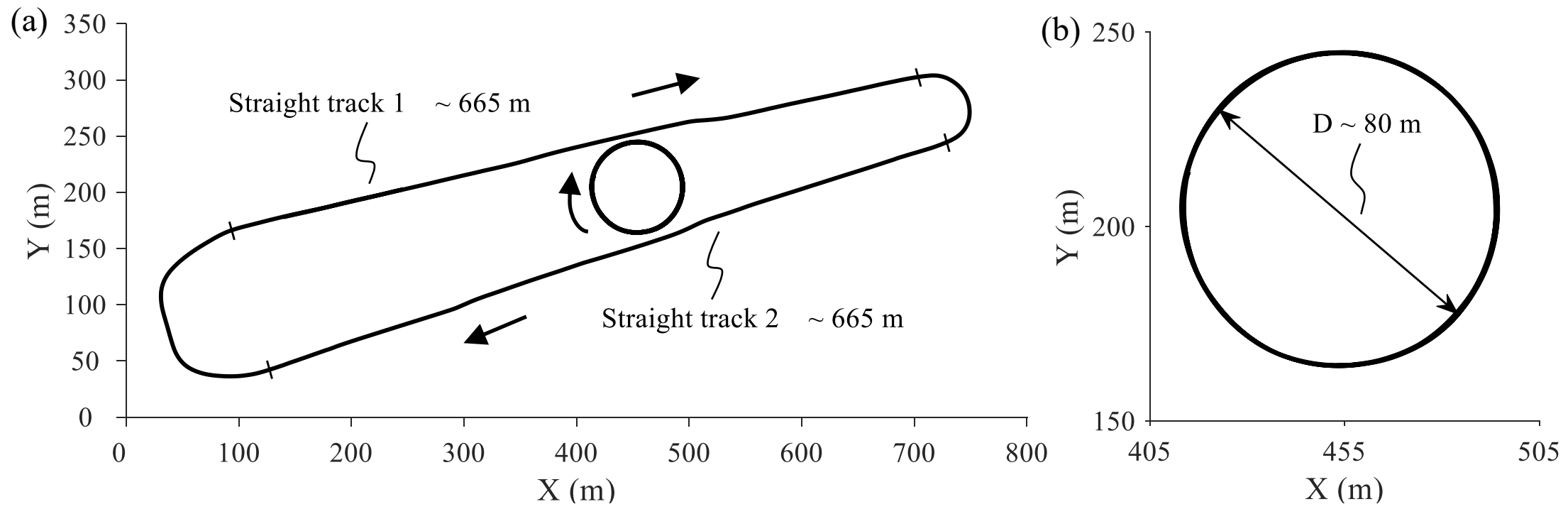

**Fig. 11.** (a) The sketch of the experimental tracks. (b) Enlarged sketch of the circular track.

The experiment used a 12-car-platoon. High-precision difference GPS devices were installed on all vehicles to record their locations and velocities every 0.1 s. The instruction to the drivers is similar to that in Experiment-II: "Drive as if you were in rush hour traffic. Follow the vehicle ahead closely, whenever safety permits".

The platoon starts from the middle of straight track 2. The leading car has the cruise control system and moves with constant speed 20 km/h on the automobile test track. After about 9 minutes, the platoon goes from the middle of straight track 1 into the circular track and moves for another 9 minutes. Then the experiment stops. Fig. 12 shows speed time series of all cars, x-coordinate of leading car, and evolution of the spatiotemporal pattern of car speed.

To compare the car-following behavior, we extract the intervals on the straight track (interval 1 & 2) and on the circular track (interval 3), as indicated by the paired oblique black dotted lines in Fig. 12(c). Fig. 13 shows the standard deviation of speeds and accelerations of all cars. Fig. 14 shows the average distance headway between cars. One can see that the data on the straight track and on the circular track are in good agreement with each other.

We test the similarity of speed/acceleration standard deviation between circular and straight tracks, respectively. The two paired data of speed/acceleration standard deviation shown in Table 2 are used in the *t*-test and their difference was used in the Mann-Kendall trend test. The *p*-value in the *t*-test equals 0.0393 and 0.0216 for the two pairs of data in Fig. 13, respectively. The null hypothesis that there is no significant difference in the speed/acceleration standard deviation between circular and straight tracks could not be rejected ($p$-value > 0.01). The *p*-value in the Mann-Kendall trend test equals 0.0164 and 0.1499, respectively. The null hypothesis that no significant trend exists could not be rejected ($p$-value > 0.01). The two tests thus indicate that driving on the straight track and on the circular track can be considered equivalent.

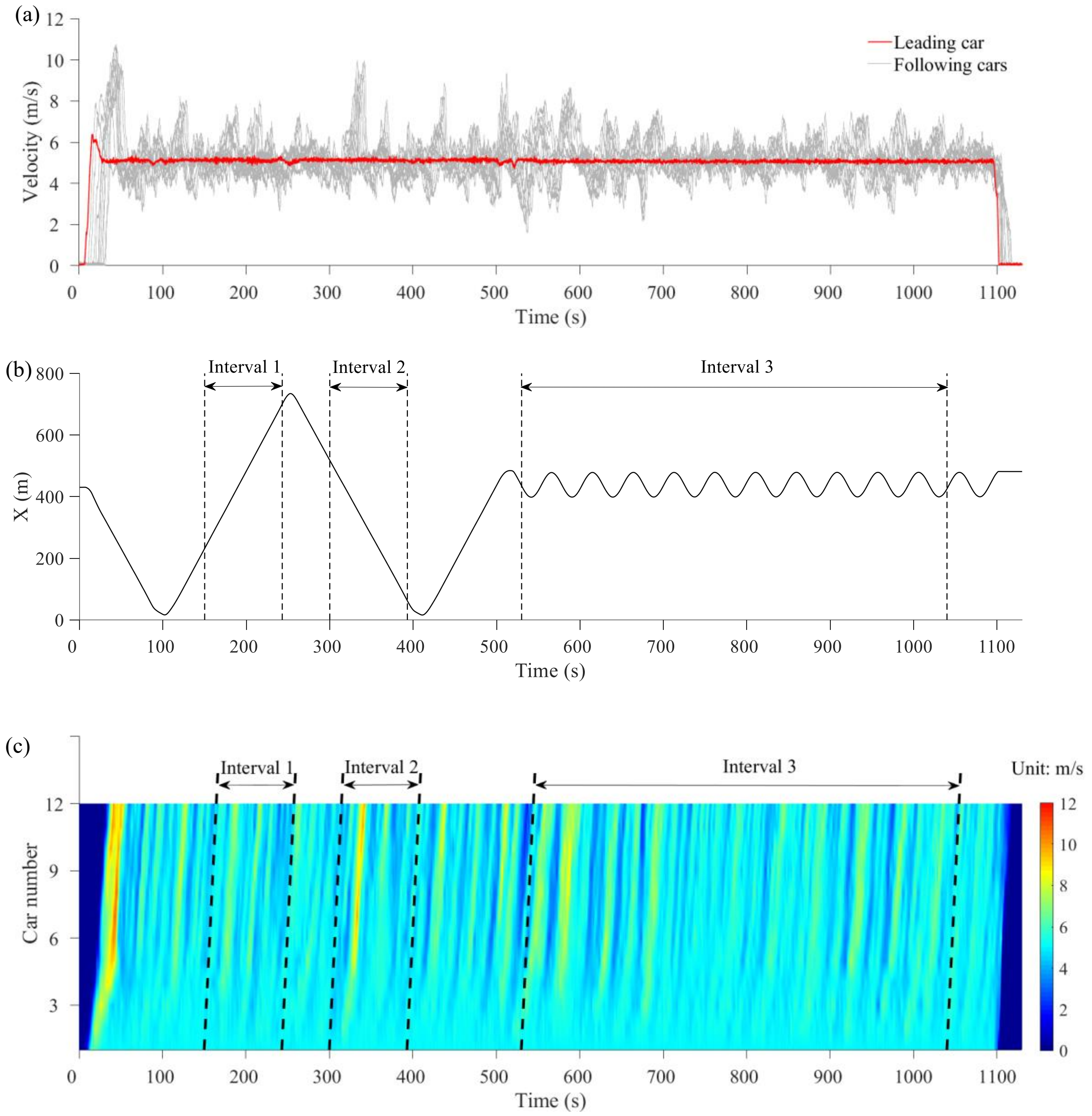

**Fig. 12.** (a) The velocity profile of all cars, (b) the x-coordinate of leading car, and (c) evolution of the spatiotemporal pattern of car speed in Experiment-III.

Moreover, Fig. 13 also compares the Experiment-III data with the Experiment-I data and the Experiment-II data. Fig. 13(a) shows that the growth trend of speed standard deviation in Experiment-III is consistent with that in Experiment-I and Experiment-II. However, the acceleration standard deviation in Experiment-III is larger than the Experiment-I data, and its growth trend is consistent with the Experiment-II data, see Fig. 13(b). We compare the average distance headway of all cars in Fig. 14. One can see that the results are close between the Experiment-II data and the Experiment-III data. This further confirms that, the reason why acceleration standard deviation is larger in Experiment-II and Experiment-III is that drivers were asked to follow closely. In contrast, in Experiment-I, drivers were asked to drive normally instead of closely.

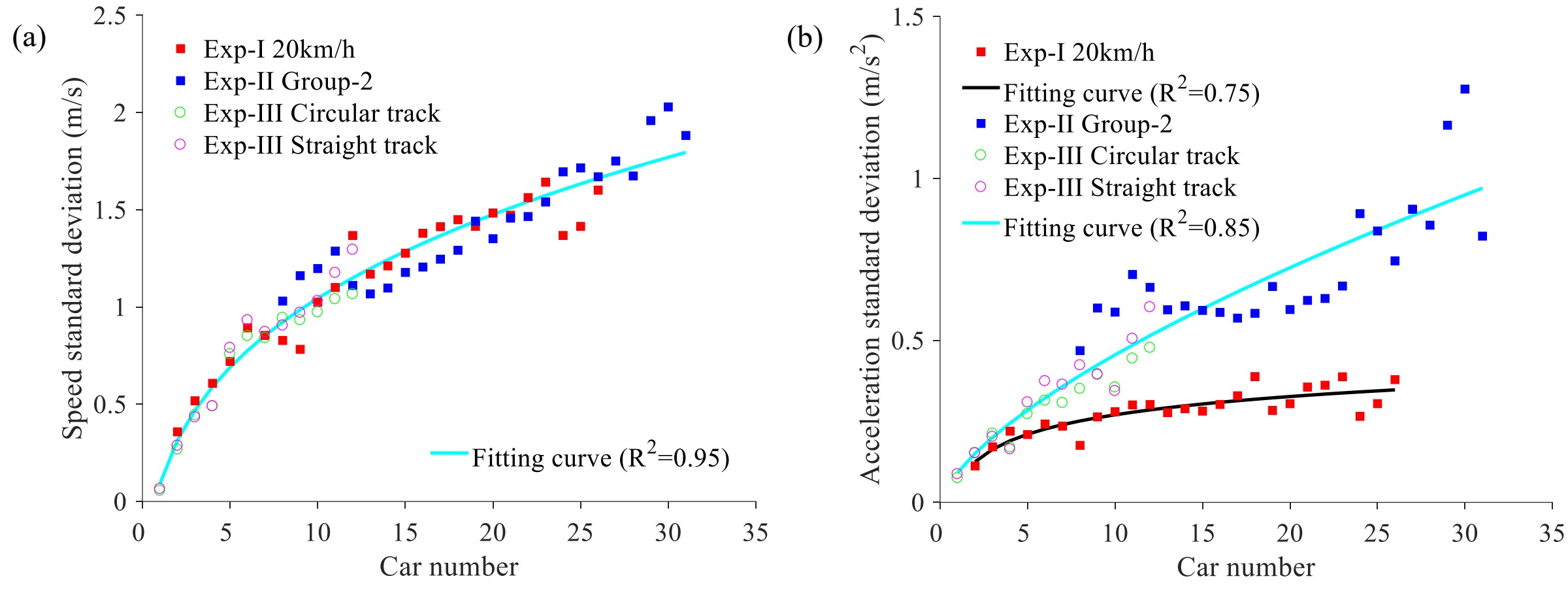

**Fig. 13.** Comparison of (a) the speed standard deviation and (b) the acceleration standard deviation of all vehicles among Experiment-I, Experiment-II and Experiment-III at the similar speed of the leading car. Note that the Experiment-II data and the Experiment-III data have been shifted horizontally. The fitting curve is given by $f(x) = ax^b + c$, where (a) $a$ = 1.0440, $b$ = 0.2820, $c$ = -0.9531; (b) $a$ = 0.1024, $b$ = 0.6579, $c$ = -0.0104 (the cyan line); $a$ = -1.1800, $b$ = -0.0869, $c$ = 1.2370 (the black line).

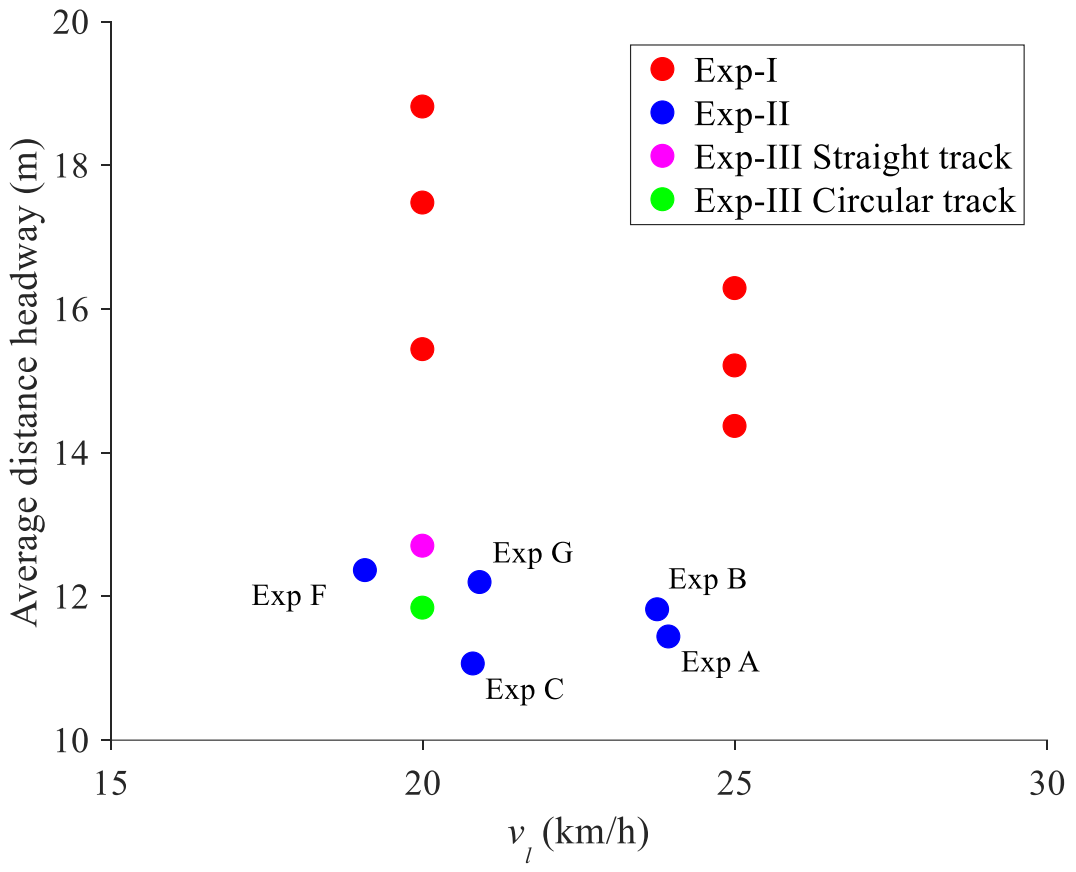

**Fig. 14.** Comparison of the average distance headway among Experiment-I, Experiment-II and Experiment-III. Each data point corresponds to one run of experiment. $v_l$ denotes the average velocity of the leading car.

## 6. Conclusions

To summarize, this paper makes a comparison analysis on the oscillation features of traffic flow between the 25-car-platoon experiment (Experiment-I), the car-following experiments in the USA (Experiment-II) and the 12-car-platoon experiment in China (Experiment-III). Our analysis reveals that (i) the set of experiment on the circular track can be considered equivalent to that on the straight track (Fig. 13); (ii) the speed standard deviation exhibits concave growth characteristic in both experimental instructions ("catch up" vs. "drive normally"; Fig. 7); (iii) the growth pattern of acceleration standard deviation is remarkably different (Fig. 9). To further study the phenomena, we compare the average distance headway in Experiment-I, Experiment-II, and Experiment-III (Fig. 10 and Fig. 14). The finding reveals that the acceleration standard deviation is related to the distance headway, which can be influenced by the experimental instruction. This finding, we believe, has implications in the future experimental study on traffic flow dynamics. In the future work, a more comprehensive comparison study will be performed concerning longer platoon, larger speed and more diversified experimental instructions.

## Acknowledgment

This work is supported by National Key R&D Program of China (No. 2018YFB1600900), the National Natural Science Foundation of China (Grants No. 71621001, 71931002), and Beijing Natural Science Foundation (Grant No. 9172013).